**Reconfigurable Plasmonic Chirality: Fundamentals and Applications**

*Frank Neubrech, Mario Hentschel, and Na Liu\**

Dr. F. Neubrech, Prof. Dr. N. Liu
Kirchhoff-Institute for Physics, Heidelberg University, Im Neuenheimer Feld 227, 69120 Heidelberg, Germany
Max-Planck-Institute for Intelligent Systems, Heissenbergstraße 3, 70569 Stuttgart, Germany
E-mail: na.liu@kip.uni-heidelberg.de

Dr. M. Hentschel
4th Physics Institute and Research Center SCoPE, University of Stuttgart, Pfaffenwaldring 57, 70569 Stuttgart, Germany
Center for Integrated Quantum Science and Technology (IQST), Pfaffenwaldring 57, 70569 Stuttgart, Germany

**Abstract**

Molecular chirality is a geometric property, which is of great importance in chemistry, biology, and medicine. Recently, plasmonic nanostructures that exhibit distinct chiroptical responses have attracted tremendous interest, given their ability to emulate the properties of chiral molecules with tailored and pronounced optical characteristics. However, the optical chirality of such man-made structures is in general static and cannot be manipulated post-fabrication. In this research news, different concepts to reconfigure the chiroptical responses of plasmonic nano- and microobjects are outlined. Depending on the utilized strategies and stimuli, the chiroptical signature, the three-dimensional structural conformation or both can be reconfigured. Optical devices based on plasmonic nanostructures with reconfigurable chirality feature a great potential in practical applications, ranging from polarization conversion elements to enantioselective analysis, chiral sensing, and catalysis.

# 1. Introduction

Plasmonic chirality has been an active field of research in the past few years.[1–3] In light of the numerous experimental and theoretical studies, it is evident that this lively field has been rigorously advanced forward with ever growing new possibilities and ideas fueled by the fascination for chirality in general. Very recently, this field has been steered to a fascinating direction, coined reconfigurable plasmonic chirality.

The words, "reconfigurable" and "plasmonics" are well known. However, a precise definition of the terminology "reconfigurable plasmonic chirality" appears elusive. Such a definition can be, in fact, less clear-cut than one would expect. In order to elucidate this, let us consider two extreme cases. First, we take an example of an elongated plasmonic spiral, either composed of individual metallic constitutes or made of solid metal. It can exhibit notably different chiroptical spectra under light illumination parallel or perpendicular to its long axis.[4] It has been demonstrated that its far-field chiroptical responses can even invert, meaning the "spectral chirality" can change without the geometry of the structure itself undergoing any changes.[5] Second, we take a number of differently sized plasmonic spheres, which can be freely arranged, e.g., by opto-thermoelectric fields.[6] Here one can, on a single structure level, freely rearrange the geometry and thus fully flip the handedness of the structure, giving rise to inverted far-field chiroptical responses as well.

Following up on the definition: which of these two cases is "reconfigurable plasmonic chirality"? We could in fact reckon several options, i.e., the chiroptical far-field signals, the mere geometric properties, or both. The answer will depend on the applications one has in mind. For instance, if the polarization state of the far-field radiation is to be manipulated, the near-fields or nanoscale processes that enable this manipulation become less crucial. If, however, the processes of interest occur in the near-fields, as for any type of plasmonic sensing,[7] the properties of the near-fields and the nature of the plasmonic modes are of utmost importance.

The outline of this research news is as follows: we will first introduce a theoretical model to understand and interpret the chiroptical features of plasmonic structures. Then, we will discuss different reconfiguration strategies. We will particularly focus on those geometries, which allow good control over structural chirality as well as optical chirality. In the end, we will show that these structures are promising for many of the perceived and envisioned applications and are in fact even mass producible.

**2. Modeling and Interpreting the Chiroptical Responses**

In experiments, it is often found that the interpretation of chiral spectra can be challenging. The characteristic spectral features in chiral responses basically stem from the electronic and optical properties of the chiral medium.[8] In molecular systems, these spectral signatures are a consequence of the electronic transitions. As chirality is an intrinsic three-dimensional property, these signatures are uniquely correlated to the conformation of the system under investigation.

Compared to molecular chirality, plasmonic chirality is a new field, sharing similarities but showing evident differences.[3] Such plasmonic systems can achieve strong circular dichroism (CD), i.e., difference in absorption ($A$) of left-handed (LH) and right-handed (RH) circularly polarized light. Different from natural molecules, plasmonic building blocks are fairly large objects with significant dipole strength. In addition, with the advent of sophisticated top-down and bottom-up techniques, plasmonic nano- and microobjects can be structured, shaped, and assembled nearly at will, rendering plasmonic chirality tailorable.[9,10] Consequently, many different routes to constructing chiral plasmonic nanostructures have been pursued. The so-called planar chiral structures generally exhibit weak chiroptical responses.[1] In contrast, three-dimensional chiral arrangements of metallic nanoparticles or solid metallic spirals may possess much larger chiroptical responses.[11–13] The difficulty in understanding plasmonic chirality lies in the task to attribute a multitude of excited plasmonic modes to the observed spectral signatures. Apart from the fundamental dipole modes, there are also higher-order ones, which are optically excitable. Furthermore, plasmonic modes are often spectrally broad and thus tend to overlap.

A number of intuitive models and numerical methods have been created to gain insight into optical chirality.[14–17] One of the important approaches is the so-called Born-Kuhn model.[18,19] In this model, CD is interpreted as a consequence of two coupled harmonic oscillators. It is uniquely suited to understand chiral molecules as well as plasmonic structures, because both electronic dipoles and plasmonic dipoles can be represented by harmonic oscillators.

The equations of motion for the coupled harmonic oscillators with loss $\gamma$, eigenfrequency $\omega_0$ (degenerate oscillators), coupling strength $\xi$, mass $m$, charge $e$, center position position $z_0$, and relative distance $d$ read

$$\ddot{x}(t) + \gamma\dot{x}(t) + \omega_0^2 x(t) + \xi y(t) = -\frac{e}{m}E_x e^{-i\omega t + ik(z_0 + d/2)} + c.c. \tag{1}$$

$$\ddot{y}(t) + \gamma\dot{y}(t) + \omega_0^2 y(t) + \xi x(t) = -\frac{e}{m}E_x e^{-i\omega t + ik(z_0 - d/2)} + c.c. \tag{2}$$

with $c.c.$ denoting the complex conjugate.[20] The arrangement of the two coupled oscillators is sketched in the inset of **Figure 1**a.

By solving these equations, the optical constants of the system can be extracted.[21] The measurable quantities in such a system include optical rotatory dispersion (ORD) and CD. It is important to note that ORD ($\theta$) and CD are Kramers-Kroning-related and therefore contain the same information. One obtains

$$\theta = \frac{\omega}{2c}Re(\rho) \tag{3}$$

and

$$A_{LH} - A_{RH} = 2\frac{\omega}{c}Im(\rho) \tag{4}$$

where the so-called nonlocality parameter $\rho$ is defined as

$$\rho = \frac{dN_0 e^2}{3m} \frac{\xi}{(\omega_0^2 - i\gamma\omega - \omega^2)^2 - \xi^2}. \tag{5}$$

As shown in Figure 1a, the CD spectrum exhibits a characteristic bisignate profile, whereas ORD reaches the maximum strength at the zero-crossing of the CD response.

Apart from the mathematical expressions discussed above, one can also understand the underlying physics in an intuitive way.[21] Due to coupling, two collective modes are formed. One is a symmetric mode with an in-phase oscillation of the two dipoles and the other is an antisymmetric mode with an out-of-phase oscillation of the dipoles as illustrated in Figure 1b. Each of these two modes has a distinct sense of rotation, which means both modes have an intrinsic chiral nature. They can be predominantly excited by left-handed circularly polarized and right-handed circularly polarized light, respectively. In the most general case, the two collective modes are excited simultaneously. Therefore, both modes can appear as spectral signatures in the RH and LH transmission (absorption, reflection) spectra. This explains in part the complicated spectra generally observed in chiral plasmonic systems. In addition, higher-order modes are also excitable and are often spectrally overlapping.

## 3. Substrate-Based Reconfiguration Strategies

The studies on plasmonic chirality have been restricted to static chiral systems for a long time. Plasmonic chirality tuning has mainly been achieved by fabricating a series of static samples, in which the geometric parameters, such as the lengths, shapes, relative distances of the individual components or the surrounding media of the structures were varied successively.[3,22–25] Integration of phase-change materials with chiral plasmonic structures has enabled reconfiguration of the chiroptical responses without changing the structural conformation. For example, Yin and co-workers incorporated a thermal-responsive material, germanium-antimony-tellurium (GST), as spacer to tune the chiroptical responses of a plasmonic system comprising two orthogonally stacked gold bars (see **Figure 2**a).[26] Upon heating to 180°C, GST transited from the crystalline phase to the amorphous phase. This transition was accompanied with a large refractive index change. As a result, the CD spectrum could be shifted to longer infrared wavelengths. In addition, the authors demonstrated a change of the CD spectrum by superimposing the responses from two chiral units of opposite handedness (one with GST and one without). By heating the sample, the CD of one unit was detuned, leading to a signal change of the overall CD spectrum without changing the structural configuration. Such a scheme for reconfiguration of the chiroptical responses based on the refractive index control of the surrounding media will be useful in realization of optical devices with switchable polarizations.

The chiroptical responses of micro- and nanostructures can also be reconfigured by manipulation of their intrinsic electronic properties.[27–29] Zhang and co-workers demonstrated reversible photoinduced CD tuning without changing the structural configuration of the plasmonic system.[27] Using multilayer lithography, they fabricated a sophisticated microstructure consisting of two virtual chiral units (called meta-atoms) of opposite handedness as shown in Figure 2b. Each unit was a three-dimensional split resonator. The structural configuration of the meta-atoms was the same but the arrangements of gold and silicon (Si) sections in the two units differed. In the absence of photoexcitation, the silicon pad broke the mirror symmetry of the structure. This caused strong optical chirality at CD resonance originating from the coupling between electric and magnetic fields. Photocarriers were generated in the Si pads upon illumination by a near-infrared laser and Si became conductive in the THz spectral range. Consequently, the plasmonic response of each unit was changed and the sign of the overall CD spectrum flipped. Obviously, the chiroptical responses of the meta-molecules were reconfigured without changing the structural conformation.

Structural reconfiguration, however, has been demonstrated using strain and pressure as stimuli.[30–32] Kan and co-workers, for example, fabricated freestanding Archimedean spirals made of gold on a silicon membrane as shown in Figure 2c.[32] In the initial state, the structure was achiral and no CD was observed. When the membrane was placed in an appropriately designed cell, pressure could be applied from either side of the spiral. Due to the finite flexibility of the spiral, it could be elongated to the adjacent volume, when pressure was applied on the opposite side. Consequently, the planar spiral was changed to a three-dimensional chiral structure. As the pressure difference between the two sides of the membrane could be freely controlled, the structure could be continuously transformed between the maximum extruded LH and RH states. This transition was optically visualized as continuously tunable and well-modulated CD spectral changes in the THz spectral range. It is noteworthy that this platform allowed control over the geometrical handedness of the spiral as well as over the amplitude of the CD spectra. The plasmonic modes generated in the structure could be interpreted as standing wave type excitations extending over the entire length of the spiral. The spectral signatures were thus of different origin compared to those in the case of the Born-Kuhn structure and can be understood in analogy to the Condon model,[33] where an electron is considered to oscillate in a chiral potential.

While this work is impressive, the presented structure is still not fully reconfigurable. The geometrical parameters, such as the length, pitch, and diameter of the spiral are fixed during fabrication and cannot be manipulated afterwards. Alternatively, Lin and co-workers arranged individual colloidal metallic and dielectric nanoparticles on a substrate surface via opto-thermoelectric fields generated by a focused laser beam.[6] This allowed to assemble and disassemble a cluster composed of three differently sized nanoparticles into chiral geometries with intended handedness as sketched in Figure 2d. Rearrangement of the particles could transform the cluster into its enantiomer as well as to varied geometries of the same or opposite handedness. The corresponding chiral scattering spectra featured highly complex modes. In such a cluster, the dipolar resonances excited in the individual nanoparticles mixed and hybridized to form collective modes of chiral nature. Each particle supported three energetically degenerate dipolar resonances that were coupled, forming a large number of spectral features. The mode formation could be described and understood in terms of coupled harmonic oscillators, similar to the case of the Born-Kuhn model. However, the rather large number of oscillators makes the modeling process challenging.

## 4. Solution-Based Reconfiguration Strategies

The aforementioned substrate-based reconfiguration strategies hold great potential for applications. Due to the smart superposition designs, they are ideally suited to manipulate far-field optical properties of light, such as polarization state or light propagation as also discussed in the introduction section. However, these approaches may have restrictions due to the need of substrates and limited throughput for practical applications. Hybrid chiral nanostructures assembled using bottom-up approaches, such as DNA nanotechnology, provide an elegant solution.[34] Such solution-based reconfiguration concepts may be favorable for chiral catalysis or sensing applications in biological and chemical environments, where interactions between analytes and chiral plasmonic near-fields are indispensable.

**Figure 3**a depicts a reconfigurable chiral structure assembled using the DNA origami technique.[35] The process involves the folding of long DNA scaffold strands by hundreds of shorter staple strands into predefined shapes. The so-called DNA origami can be dressed with a variety of nanoscale objects including metallic particles, quantum dots, proteins, among others using capture strands. The structure shown in Figure 3a consisted of two linked DNA origami bundles, which were functionalized with two gold nanorods, forming a three-dimensional plasmonic cross. The configuration of the plasmonic cross could be modulated by addition of the corresponding DNA fuel strands and the structure was switched among LH, relaxed, and RH states. The switching of the structure could be monitored using CD spectroscopy in real time.

If we compare the chiral plasmonic structures assembled by DNA origami with those fabricated on substrates, the degree of reconfiguration of the former is much higher and applies to the geometrical handedness as well as to the chiroptical properties. One of the key benefits lies in the fact that DNA nanotechnology allows for solution-based fabrication of very large quantities, an important concern in any practical application. Additionally, the assembled structures are freely dispersed in solution. Undesired effects induced by interfaces or relative orientations between the structures and the incident light fields can be efficiently suppressed.

Another benefit afforded by DNA nanotechnology is its unique capability to reconfigure structures with different stimuli including DNA,[35–42] RNA,[43,44] aptamers,[45,46] light,[47–50] pH,[48,51,52] temperature,[46,53] among others.[5,54,55] Figure 3b shows one of the examples, in which light is adopted as stimulus.[50] The structure consisted of two crossed DNA bundles with two gold rods attached. The reconfiguration of the structure was enabled by a DNA lock modified by azobenzene molecules, which can undergo *trans-cis* photoisomerization. More specifically, illumination with ultraviolet (UV) light led to the dehybridization of the DNA lock and the plasmonic cross was transformed to the relaxed state. Illumination with visible light led to the closing of the DNA lock and thus transformed the metamolecule back to the chiral state. In this case, the trigger for the reconfiguration was light, which is a fascinating prospect per se and has been demonstrated also in other works.[47–49] Liu and co-workers took one additional step and incorporated the metamolecules into liquid crystals. The random orientations of the metamolecules were preserved, because the liquid crystals added a rigid and stable support to the system. The authors demonstrated that the metamolecules fully retained their switchability in spite of the support. The overall compound-structure is thus highly interesting for applications, as integration of liquid crystal-based materials is a standard procedure in today's technology. For instance, thin polymer-like films could be utilized as a means for polarization control and manipulation.

An example of the chiral plasmonic structures that can respond to multiple stimuli is presented in Figure 3c. Here, Jiang and co-workers utilized two rhombus DNA origami sheets to assemble L-shaped chiral dimers.[48] The seam between the two sheets was connected using different linkers, which had two functions. First, predesigned "zippers", such as the disulfide bonds, could be incorporated. The dimers were disassembled upon contact with glutathione. As a result, the chiroptical responses vanished. Second, appropriately designed DNA strands could be elongated or contracted upon exposure to certain stimuli, leading to the modification of the CD strength. The authors realized a pH-responsive system by incorporation of i-motif sequences into the linker strands. Specially designed complementary strands formed a duplex with the i-motif keeping the nanorods apart at neutral pH. Upon acidification of the system, the complementary strands were released and the interrod distance decreased. Photoregulation of the chiral plasmonic system was achieved by incorporating telomere DNA sequences into the linker strands in combination with aforementioned azobenzene moieties. Upon irradiation with visible light, the linker formed a folded G-quadruplex stabilized by the *trans* form of the azobenzene derivatives. Illumination with UV stimulated the formation of the *cis* state and the folded G-quadruplex became stretched, resulting in a larger interrod distance. The structural design was carefully optimized so that the two origami sheets allowed for a corner-stacked configuration. The coupling strength was thus maximum, leading to the largest spectral splitting of the symmetric and antisymmetric modes possible in this configuration. Similar to the Born-Kuhn case discussed in Figure 1, a pronounced bisignate CD profile was achieved as shown in Figure 3c. Such a versatile and optimized platform responding to multiple stimuli is very promising for the detection of minute amounts of analytes.[56]

Man and co-workers demonstrated autonomous tuning of chiroptical responses using pH-sensitive triplex DNA segments in tetrahedral DNA origami templates decorated with gold nanorods.[52] The DNA triplex segment consisted of a duplex located in the tetrahedral DNA origami and a thiolated single stranded DNA modified on gold nanorods. For this distinct configuration, pH-values below (above) a critical value of 7.2 initiated the assembly (disassembly) of chiral nanostructures. The authors coupled such a proton-responsive chiral system to a positive-feedback chemical reaction network (CRN), which provided alternating pH-values. By adding urea (fuel) to the network, the pH-value was reduced below the critical level and thus, the assembly of the chiral structures was triggered. An enzyme, also present in the CRN, catalyzed the hydrolysis of urea and released $NH_3$ causing a pH increase. At the critical pH level, the disassembly was initiated. The graphs in the lower half of Figure 3d depicts the CD responses of the solution at 708 nm wavelength upon cyclic pH changes. It is noteworthy that the pH stimulus has also been used to reconfigure other DNA-based chiral plasmonic nanosystems.[48,51]

A final example of a multi-stimulus structure responsive to aptamers and temperature is shown in Figure 3e. Here, Zhou and co-workers have assembled two gold nanorods onto a DNA bundle and a DNA plate, respectively.[46] The rotary bundle was connected to the rectangular sheet at its center via the scaffold DNA strand. Additionally, pairs of split adenosine triphosphate (ATP) and cocaine (COC) aptamers were positioned on the DNA sheet along the a-a and b-b diagonal direction, respectively. Each split aptamer consisted of two DNA strands. One was extended from the rotary bundle and the other one was from the origami sheet. Before target binding of ATP and COC aptamers, the plasmonic system exhibited a RH preference at 35°C (black curve). Upon addition of the ATP molecules, the chiroptical responses were flipped to the LH state (red curve). On the contrary, when COC molecules were added, the plasmonic nanostructures were driven more effectively to the RH state, showing stronger RH chiroptical responses (blue curve). As expected for stacked rod structures of Born-Kuhn type, well pronounced and clean bisignate responses were observed in the CD spectra shown in Figure 3e. The authors further demonstrated that the handedness of the chiral plasmonic system can be changed by temperature as well.

The presented examples demonstrate the capability of reconfigurable plasmonic chirality applications. Their reconfigurable nature enables the production of reusable DNA-based chiral sensors as well as fast switchable light manipulation devices. The reconfigurable chiral subunits could be straightforwardly implemented in already existing optical elements by lithographic processes. On the other hand, DNA-based reconfigurable structures in solution are promising for injection in cells, chiral catalysis and in vitro sensing of minute amounts of biological and chemical analytes.

## 5. Conclusions

Significant efforts have been invested in order to understand, tailor, and manipulate the chiroptical responses of plasmonic nano- and microstructures. Reconfiguration can apply to mere structural or spectral properties as well as to both. A change in the chiroptical response, for instance, does not necessarily require a reconfiguration of the geometry itself. Similarly, an inversion of the geometrical handedness does not necessarily result in an inversion of the chiroptical response. Depending on the aimed application, it is thus crucial to choose the best suited geometry. In applications requiring the interaction with chiral near-fields, a mere change in the far-field properties due to, for example, a smart superposition design of several subunits is insufficient. On the other hand, if the far-field properties are to be manipulated, the nanoscale reconfiguration mechanism is of minor interest.

A rich variety of structures, ansatzes, concepts, and designs have been proposed and studied. At this point, it is indeed possible to select the ideally suited concept and tailor it for applications. Self-assembled DNA nanostructures are for example uniquely tailorable to multiple and vastly different stimuli, making them prime candidates for nanoscale sensors. Their signal-to-noise ratio and sensitivity are outstanding. Top-down fabricated structures with smart subunit designs, on the other hand, are ideal for far-field manipulation. In these cases, bulk properties, such as the refractive index of the surrounding medium, are being targeted for reconfiguration. While this is not a structural handedness reconfiguration, it has the benefit of no "wear and tear", as nanoscale rearrangement is not required. We believe that the field of reconfigurable plasmonic chirality has just entered the application phase. Apart from ultrasensitive nanoscale sensors, polarization control in micro- and nanoscale integrated optical circuitry is also a very promising direction.

**Acknowledgements**

N.L. and F.N. acknowledge funding from the European Research Council (Dynamic Nano) grant. M.H. was supported by the European Research Council (Complexplas), DFG, BW Stiftung, MWK Baden-Württemberg (ZAQuant), and BMBF.

**Figures**

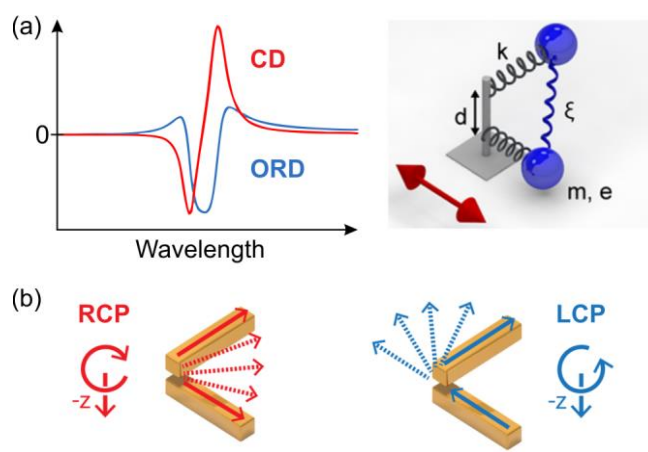

**Figure 1.** (a) Born-Kuhn model: Circular dichroism (CD) and optical rotatory dispersion (ORD) calculated from the right-handed classical coupled-oscillator model (see schematic drawing). Parameters are given in the text. (b) Modes excited by right- and left-handed circularly polarized light (RCP and LCP, respectively) for the right-handed configuration of corner-stacked nanorods spaced at an effective quarter-wavelength. Panels (a) and (b) adapted with permission.[21] Copyright 2013, American Chemical Society.

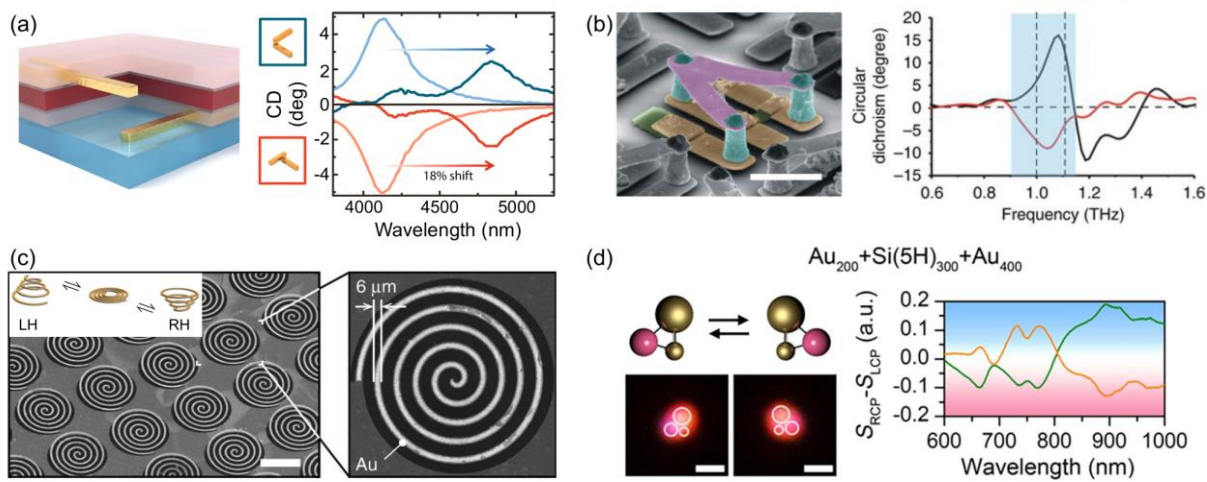

**Figure 2.** Different strategies for reconfiguration. (a) Spectral reconfiguration of a chiral plasmonic system composed of two corner-stacked, orthogonal gold nanorods separated by the phase change material germanium-antimony-tellurium (GST, dark red). Upon heating, GST changes its refractive index resulting in a shift of the CD signal. CD measurements of left- (blue) and right-handed (red) chiral structures are shown. (b) Photoinduced spectral reconfiguration. SEM image of a three-dimensional meta-molecule consisting of two chiral units determining the overall chiroptical response. By optically manipulating the charge carrier density in the silicon patches (green), the spectral handedness is reconfigured. The scale indicates 10 µm. (c) Structural reconfiguration of deformable micro-electro-mechanical (MEMS) spirals operating in the THz spectral range. Depending on an applied pressure, the structural and optical handedness transit from left-handed (LH) to right-handed (RH). The scale bar is 100 µm. (d) Structural reconfiguration of chiral meta-molecules composed of gold (Au) and silicon (Si) nanoparticles using opto-thermoelectric fields. The micrographs and CD spectra proof the left- and right-handed conformation before and after manipulation, respectively. Scale bar: 1µm. Panel (a) adapted with permission.[26] Copyright 2015, American Chemical Society. Panel (b) reproduced with permission.[27] Copyright 2012, Macmillan Publishers Limited. Panel (c) adapted under the terms of the CC-BY Creative Commons Attribution 4.0 International License.[32] Panel (d) reproduced with permission.[6] Copyright 2019, Elsevier.

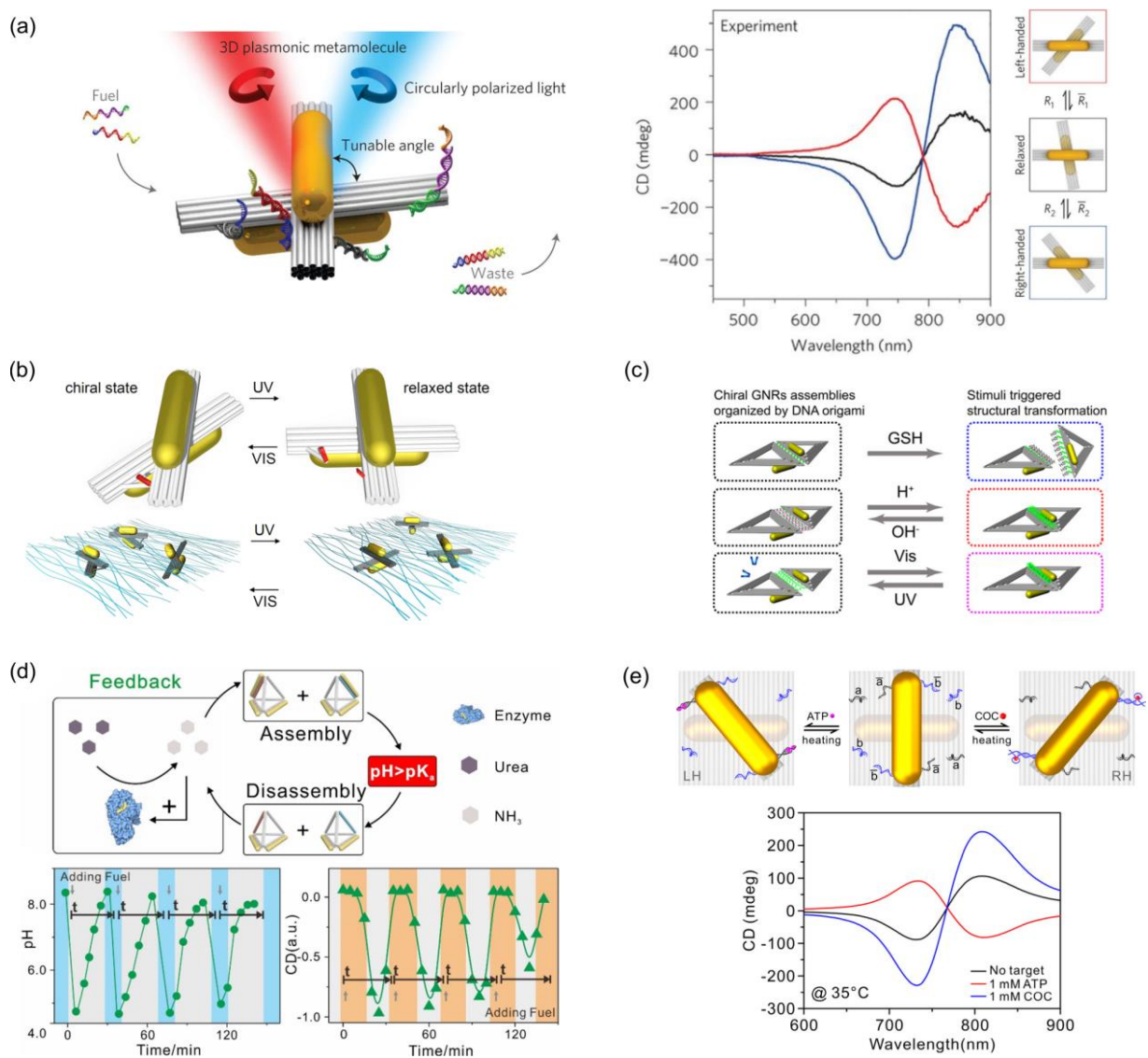

**Figure 3.** Structural reconfiguration of DNA-origami-based plasmonic chiral nanoarchitectures exposed to different stimuli. (a) Cross-like DNA bundles decorated with gold nanostructures. With specially designed DNA fuels ($R_1, R_2, \overline{R_1}, \overline{R_2}$) locks can be opened and closed reversibly providing an active control of the structural and optical handedness. Three conformational states are identified via CD spectroscopy: relaxed (black), right-handed (blue) and left-handed (red) (b) Photoinduced reconfiguration of chiral cross-like DNA bundles decorated with gold nanorods. The plasmonic system is incorporated in a liquid crystal environment. Illumination with visible (VIS) and UV radiation reversibly switches the relaxed state (achiral) and the locked state (chiral) as confirmed by CD spectroscopy. (c) Chiral nanorod assemblies organized by DNA origami. Depending on the interlock-design, the chiroptical responses can be reversibly reconfigured using light or different pH-values. Glutathione (GSH), however, introduces an irreversible change. (d) pH-induced transient reconfiguration of plasmonic meta-molecules composed of tetrahedral DNA-origami and gold nanorods. If the pH-value of the non-equilibrium system falls below a critical value (pK), handed states start to assemble. By adding a fuel, the pH value rises again and the structures disassemble after the pK is exceeded. (e) Two stimuli, temperature and aptamers, are employed to switch the handedness of a chiral plasmonic nanoarchitecture. Panel (a) reproduced with permission.[35] Copyright 2014, Macmillan Publishers Limited. Panel (b) adapted with permission.[50] Copyright 2019, The Optical Society. Panels (e), (c) and (d) adapted with permission.[46,48,52] Copyright 2017, 2019, and 2018 American Chemical Society.